\documentclass[12pt,a4paper]{article}
\usepackage{amssymb,amsmath,amscd,epsfig,bm}
\usepackage{graphicx}
\usepackage{feynmp}
\graphicspath{{Plots/}}

\title{
TeV-scale bileptons, see-saw type II and lepton flavor violation in core-collapse supernova
  }
\author{
 Oleg Lychkovskiy\thanks{e-mail: lychkovskiy@itep.ru}\hspace*{2mm} $^{\rm a}$,
  Sergei Blinnikov\hspace*{2mm}$^{\rm a,b},$
 Mikhail Vysotsky\hspace*{2mm}$^{\rm a}$
\\ ${\rm ^a}$ {\small\it Institute for Theoretical and Experimental Physics}\\
{\small\it 117218, B.Cheremushkinskaya 25,
Moscow, Russia}\\
 ${\rm ^b}$ {\small\it   IPMU, University of Tokyo,
5-1-5 Kashiwanoha, Kashiwa, 277-8568, Japan}}

\date{}

\begin{document}

\newcommand{\be}{\begin{equation}}
\newcommand{\ee}{\end{equation}}
\newcommand{\nue}{\nu_e}
\newcommand{\nul}{\nu_l}
\newcommand{\antinue}{\bar{\nu}_e}
\newcommand{\ar}{\rightarrow}
\newcommand{\rb}{\right)}
\newcommand{\lb}{\left(}
\newcommand{\kpc}{{\rm kpc}}
\newcommand{\F}{{\rm F}}

\maketitle

\begin{abstract}
Electrons and electron neutrinos in the inner core of the core-collapse supernova are highly degenerate and therefore numerous during a few seconds of explosion. In contrast, leptons of other flavors are non-degenerate and therefore relatively scarce. This is due to lepton flavor conservation.
If this conservation law is broken by some non-standard interactions, electron neutrinos are converted to muon and tau-neutrinos, and electrons -- to muons. This affects the supernova dynamics and the supernova neutrino signal. We consider lepton flavor violating interactions mediated by scalar bileptons, i.e. heavy scalars with lepton number 2. It is shown that in case of TeV-mass bileptons the electron fermi gas is equilibrated with non-electron species inside the inner supernova core at a time-scale $\sim (1-100)$ ms. In particular, a scalar triplet which generates neutrino masses through the see-saw type II mechanism is considered. It is found that supernova core is sensitive to yet unprobed values of masses and couplings of the triplet.

\end{abstract}


\newline

 \section{Introduction}

If the Standard Model (SM) is extended by a triplet of heavy scalar fields interacting with leptons and bosons of the SM, it becomes possible to generate Majorana neutrino masses through the see-saw type II mechanism \cite{Magg:1980ut}-\cite{Mohapatra:1980yp}.
Originally the mass of this triplet was assumed to be at a GUT scale; recently, however, TeV-scale see-saw became popular. In particular, it was argued that TeV-scale see-saw type II avoids electroweak hierarchy problem, encountered in GUT-scale see-saw, and meets the naturalness argument (see e.g. \cite{Xing:2009in}\cite{Abada:2007ux}). See-saw II signatures at the LHC (see e.g. \cite{Akeroyd:2007zv}\cite{Perez:2008ha}  and references therein) and in low-energy experiments (see e.g. \cite{Abada:2007ux}\cite{Akeroyd:2009nu}  and references therein) are extensively investigated.

In the present paper we point out that TeV-scale see-saw II affects the core-collapse supernova physics, namely the supernova neutrino signal and the dynamics of the explosion. This conclusion is valid in a certain area of the parameter space of the neutrino mixing matrix. The effect is due to the lepton flavor violating (LFV) four-fermion interactions, which are induced by the scalar exchange. It is known that LFV processes may change drastically the conditions in the interiors of a core-collapse supernova \cite{Mazurek}\cite{Kolb:1981mc}.\footnote{In particular, in ref. \cite{Kolb:1981mc} a Gelmini-Roncadelli model of neutrino mass generation \cite{Gelmini:1980re} was proposed as a source of LFV. It resembles  the see-saw type II model, but contains a massless Majoron, which is now excluded by the measurement of the Z-boson invisible decay width. }  Thus supernova physics appears to be an additional probe of TeV-scale see-saw II, along with LHC and low-energy experiments (such as rare lepton decays and muonium conversion).


Scalar triplet, which enters the see-saw II, is a particular case of scalar bilepton, i.e. scalar which couples to a pair of leptons. Another type of bilepton is a doubly charged singlet scalar. If possessing  TeV-scale mass, it may also be produced at LHC and influence the results of low-energy experiments. We include the analysis of its effect on the LFV in supernova in the present paper, because it mediates LFV interactions analogously to the doubly charged component of the scalar triplet.

The paper is organized as follows. In section 2 we estimate the value of effective four-fermion LFV coupling necessary to significantly affect the supernova physics. In section 3 we consider a simple extension of the SM with a doubly charged bilepton. Section 4 is devoted to see-saw type II extension of the SM. Influence of LFV inside the supernova core on supernova physics is discussed in section 5. The results are summarized in section 6.

\section{Model independent considerations}

An explosion of a non-thermonuclear supernova is a manifestations of a collapse of a progenitor star core.
Although numerous studies have failed to reproduce an explosion of a core-collapse supernova, there exists a commonly accepted general picture of the collapse (which is referred to as "standard picture" in what follows; for the details see, for example, \cite{Bethe:1990mw}).  When the mass of the iron core of a massive star reaches the Chandrasekhar limit, the infall phase of the collapse starts. The core contracts due to the gravitational attraction. Some fraction of electrons is converted to electron neutrinos through the inverse beta processes. When the density of the inner part of the core reaches the nuclear density value, $\sim 3\cdot 10^{14}$ g/cm$^3,$ the infalling matter of the outer core bounces from it. A shock wave emerges, which starts to propagate outwards. A powerful shock could expel the stellar envelope and produce the supernova explosion. However, according to the detailed numerical calculations \cite{Nadezhin:1977}, the shock wave looses its energy and stalls after some tens of milliseconds. There exist several scenarios of {\it presumably} successful explosion (e.g. \cite{BisnKog:2008}), but neither of them is justified to date by detailed self-consistent calculations with neutrino transport.

During the collapse extreme values of density, temperature, electron and electron neutrino chemical potentials are reached. For the estimates we use the reference values of the supernova core parameters as presented in Table \ref{SN conditions}.
 \begin{table}[h]
 \begin{center}
 $\begin{array}{|c|c|c|c|c|c|}
\hline
\rho & n_B &  ~~Y_e~~ & ~~Y_{\nu_e}~~ &  ~~Y_\mu~~ & ~~Y_{\nu_\mu}, ~Y_{\nu_\tau}~~ \\
\hline
2\cdot 10^{14}  ~ {\rm g/cm}^3&
1.2 \cdot  10^{38}  ~ {\rm cm}^{-3}&
0.30&
0.07 &
\sim 10^{-5} &
\sim 10^{-4} \\
\hline
\end{array}$

\vspace{0.2 cm}

$\begin{array}{|c|c|c|c|c|}
\hline
T & \mu_e & \mu_{\nu_e} & \mu_\mu & \mu_{\nu_\mu},\mu_{\nu_\tau} \\
\hline
10 ~ {\rm MeV}&
200 ~ {\rm MeV}&
160~ {\rm MeV} &
40~ {\rm MeV} &
0 \\
\hline
\end{array}$
\end{center}
\caption{Typical conditions in the inner supernova core ($m\lesssim0.5 M_\odot$) during the first \mbox{50 ms} after core bounce as obtained by us with the help of the open-code program \mbox{Boom \cite{Boom}.} Only SM interactions are taken into account.}
\label{SN conditions}
\end{table}
 This data is obtained by us with the help of the open-code programme \cite{Boom} (in which only SM interactions are taken into account),
being in good agreement with the results of more sophisticated calculations, see e.g. \cite{Sumiyoshi}\cite{Thompson:2002mw}.
The presented values are the mean values for the inner supernova core  ($m\lesssim0.5 M_\odot$); the central values are even greater.
Note that in the standard picture of the collapse electrons and electron neutrinos
 are plentiful and highly degenerate during the first hundreds of milliseconds. Their total number roughly equals the total number of electrons in the core before collapse, and decreases slowly due to the leakage of neutrinos from the core.
By contrast, there are very few non-electron neutrinos and muons. Non-electron neutrinos have negligible chemical potentials; they are created in pairs with their antiparticles, concentrations being proportional to $T^3$. Muons have chemical potential approximately equal to $\mu_e-\mu_{\nu_e},$ their concentration has an additional suppression factor $e^{-(m_\mu-\mu_{\mu})/T}$ (see Appendix).  Lepton flavor conservation law prevents electrons and electron neutrinos to turn into other types of leptons.

This picture is modified if there exist LFV processes beyond the SM \cite{Mazurek}\cite{Kolb:1981mc}. If electrons and electron neutrinos are converted into leptons of other flavors inside the inner core\footnote{Not to be confused with ordinary resonant neutrino flavor conversion due to the MSW-effect, which take place in the envelope. It affects the neutrino signal but do not affect the supernova dynamics.}, the dynamics of the collapse and the supernova neutrino signal is different from the SM case. The following LFV reactions could be {\it in principle} relevant for supernova:\footnote{We consider processes in which only leptons take part.}

\be
\begin{array}{rcl}
e^- e^- &  \rightarrow & \mu^- \mu^-,\\
e^-\nu_e & \rightarrow &  \mu^-\nu_\mu,\\
\nu_e\nu_e& \rightarrow &  \nu_\mu\nu_\mu,\\
\nu_e\nu_e& \rightarrow &  \nu_\tau\nu_\tau,

\end{array}
\label{LFV processes relevant}
\ee
\be
\begin{array}{rcccl}
e^- e^- & \rightarrow &  e^- \mu^-, & \\
e^-\nu_e& \rightarrow & e^-\nu_{\mu,\tau}, & \\
e^-\nu_e& \rightarrow & \mu^-\nu_{e,\tau}, &\\
\nu_e\nu_e & \rightarrow &  \nu_e\nu_{\mu,\tau},  \\
\nu_e\nu_e & \rightarrow &  \nu_\mu\nu_{\tau},
\end{array}
\label{LFV processes irrelevant}
\ee
The reactions are divided into two groups. In the first group all individual lepton flavor numbers, $L_e,L_\mu$ and $L_\tau$ are changed by two units or unchanged at all (e.g. for the first reaction $\Delta L_e = -2,~\Delta L_\mu = 2, ~\Delta L_\tau = 0$). In the second group at least some of the individual lepton flavor numbers are changed by one unit. As is discussed below, processes from the second group are tightly constrained by non-observation of rare decays and hardly can affect the supernova physics; in contrast, the processes from the first group can affect the supernova physics and at the same time fit the present experimental constraints.

Let us estimate the value of the cross sections of the LFV reactions potentially important for supernova core.
For this purpose we demand that such reactions should lead to the thermal equilibrium between $e$ and $\nu_e$ on the one hand and leptons of other flavor(-s) on the other hand at a certain timescale $\Delta t$
(note that equilibration between charged leptons and neutrinos of the same flavor is established virtually instantly due to the ordinary weak interactions). We consider two reference values for $\Delta t.$ The first one, $\Delta t_1 = 1$ ms, is characteristic for the dynamics of the supernova (e.g. the shock propagates through the core during several ms). The second one, $\Delta t_2,$ should characterize
the Kelvin-Helmholtz timescale of the cooling of a proto-neutron star (it is somewhat larger than the neutrino diffusion timescale, see section \ref{Standard collapse}),
which determines the long-term supernova neutrino signal. Although the inner core is deleptonized due to the diffusion at a timescale $\sim 5$ s \cite{Burrows:1986me}, we conservatively chose   $\Delta t_2 = 300$ ms, as at greater times the numbers presented in Table 1 are significantly changed.

 From the above-mentioned condition we find for the processes in which electrons are present in the initial state
\be
\sigma\gtrsim 1/(\Delta t n_e c) \simeq  3\cdot10^{-48} (300~{\rm ms}/\Delta t){\rm cm}^2.
\label{main condition}
\ee
If only neutrinos are involved in the process, the cross section should be
several
times greater, as the neutrino density is
several
times smaller.
The estimated value appears to be rather small compared to ordinary weak cross sections. For example,
\be
\sigma(\nu_e n \rightarrow e^- p)\simeq 2\cdot 10^{-39}~{\rm cm}^2
\label{weak cross-section}
\ee
for $160$ MeV neutrinos \cite{Strumia:2003zx}. 

Reactions (\ref{LFV processes relevant}) and (\ref{LFV processes irrelevant}) at low energies are described by effective four-fermion terms in the Lagrangian, the cross section being of order of $G_{\rm LFV}^2 \mu_{e,\nu_e}^2,$ where $\mu_{e,\nu_e}$ stands for the chemical potential of electron and electron neutrino correspondingly (every reaction is characterized, in general, by its own $G_{\rm LFV}$).
From (\ref{main condition}) one gets a rough estimate for $G_{\rm LFV}$ necessary for large LFV in supernova.  In case of electron-electron collisions it reads
\be
G_{\rm LFV}
\gtrsim
4 \cdot 10^{-4} \sqrt{300~{\rm ms}/\Delta t}~{\rm TeV}^{-2}.
\ee
In case of neutrino-electron and neutrino-neutrino collisions the bound is
somewhat larger due to the smaller neutrino density and chemical potential. Such bounds on \mbox{$G_{\rm LFV}\sim \lambda^2/M^2$} indicate the mass $M$ of intermediate heavy boson of order of several TeVs if the dimensionless coupling constant $\lambda$ is of order of $0.1$.

Are such values of $G_{\rm LFV}$ compatible with present experimental constraints? This question is accurately treated in the following two sections, here we provide a preliminary discussion. The answer is different for two groups of reactions, (\ref{LFV processes relevant}) and (\ref{LFV processes irrelevant}). In the former case, the best experimental limit stems from the non-observation of oscillations of muonium atom, $\mu^-e^+\leftrightarrow \mu^+e^-$\cite{Willmann:1998gd}, and reads $G_{\rm LFV} \lesssim  10^{-1}~ {\rm TeV}^{-2}.$
Although muonium oscillations involve only charged leptons, due to the SU(2) symmetry the bound on the effective constant is valid for the first three processes in (\ref{LFV processes relevant}), including those which involve neutrinos. As for the fourth process in  (\ref{LFV processes relevant}), it evades even this not very restrictive bound. Thus there is a room for LFV in supernova due to four reactions (\ref{LFV processes relevant}) with $|\Delta L_{e,\mu,\tau}|=0,2.$

As for reactions  with $|\Delta L_l|=1$ (\ref{LFV processes irrelevant}), they are tightly constrained by the rare lepton decays:
\be
\begin{array}{rcl}
\mu & \rightarrow & eee,\\
\mu & \rightarrow & e\gamma,\\
\tau & \rightarrow & lll.
\end{array}
\label{rare decays}
\ee
The constraints span from $G_{\rm LFV}\lesssim 10^{-2}~ {\rm TeV}^{-2}$ (LFV decays of $\tau$) to $G_{\rm LFV}\lesssim 10^{-5}~ {\rm TeV}^{-2}$ ($\mu^-  \rightarrow  e^-e^-e^+$), see \cite{Amsler:2008zzb} and references therein. Therefore processes (\ref{LFV processes irrelevant}) can fit condition (\ref{main condition}) only marginally or can not fit it at all. Thus they are likely to be irrelevant for LFV in supernova.

To conclude this section, we are interested in such an extension of the SM, in which four-fermion LFV processes with $|\Delta L_l|=1$ are suppressed compared to the processes with $|\Delta L_l|=2.$ In the next section we give a simple but instructive example of such extension. In section 4 we show that see-saw type II model fits this requirements for certain values of neutrino masses and mixing parameters.

\section{Doubly charged scalar singlet}

\begin{figure}[t]
\centerline{\includegraphics{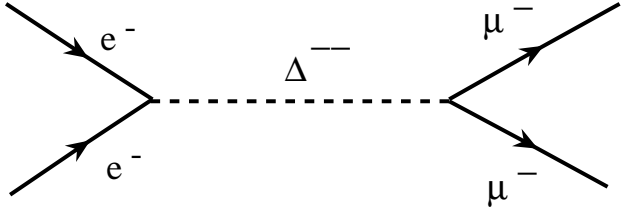}}
\caption{$ee\rightarrow \mu\mu$ LFV transition mediated by the doubly charged bilepton $\Delta^{--}$ }
\label{ee->mumu}
\end{figure}

Consider the SM extended by one doubly charged scalar $\Delta$, singlet under SU(2). It can not be coupled to quarks due to its electric charge. The only possible coupling of such scalar to leptons reads
\be\label{singlet scalar}
{\cal L}_{\Delta}=\sum_{l,l'}\lambda_{ll'}\overline{l_R^{c}} l_R' \Delta + h.c.
\ee
Here\\
$l$ and $l'$ are charged leptons, $l,l'=e,\mu,\tau;$\\
$\lambda_{ll'}$ is a matrix of coupling constants, summation over lepton flavors is implied.\\
C-conjugated right-handed fermions here and in what follows are defined as $\psi_R^{c}\equiv(\psi_R)^c =\gamma^2 (\psi_R)^*,$ standard representation of gamma-matrices being used.

Let us choose the coupling matrix to be proportional to unit matrix: $\lambda_{ll'}=\lambda \delta_{l l'}.$
In this model low-energy processes with $|\Delta L_{l}|=1$ are forbidden.\footnote{Note that in general the coupling matrix $\Lambda\equiv||\lambda_{ll'}||$ is not invariant under rotations of the charged lepton basis. It transforms according to $\Lambda'= U^T \Lambda U=\lambda U^T U,$ where $U$ is a matrix of a  unitary rotation, $U^\dagger U=1$. However, in case when $U$ does not contain complex phases one gets $U^\dagger=U^T$ and $\Lambda'=\Lambda.$ The role of complex phases is investigated in more detail in the next section.}

The exchange of doubly charged bilepton gives rise only to one LFV reaction, interesting in the context of supernova core conditions (see Fig. \ref{ee->mumu}):
\be
e^-e^-\rightarrow \mu^-\mu^-.
\ee
The corresponding effective low-energy LFV term reads
\be
{\cal L}_{LFV}^{eff} = \frac{\lambda^2}{M_\Delta^2}\overline{e_R^c} e_R(\overline{\mu_R^c} \mu_R)^\dagger,
\ee
which gives (see e.g. \cite{Cuypers:1996ia}\cite{Maalampi:1993ju}\cite{Raidal:1997tb})
\be
\sigma(ee\rightarrow \mu\mu)=\frac{\lambda^4}{M_\Delta^4}(1-2m_\mu^2/s)|t_{max}-t_{min}|/8\pi.
\label{sigma}
\ee
Note that one should take into account the combinatoric factors and the identity of fermions in the final state while calculating this cross section.
In the center-of-mass reference frame  $|t_{max}-t_{min}|=4E_e \sqrt{E_e^2-m_\mu^2}.$  Taking $E_e\simeq\mu_e \sim 2m_\mu$ one obtains
\be
\sigma(ee\rightarrow \mu\mu)\simeq2\cdot 10^{-42} \left( \frac{\lambda^4}{(M_\Delta/{\rm TeV})^4} \right) ~{\rm cm}^2.
\label{sigma}
\ee
Comparing this with the estimate (\ref{main condition}) one finds that the supernova physics is modified if
\be
\frac{\lambda^2}{(M_\Delta/{\rm TeV})^2} \gtrsim 10^{-3} \sqrt{300~{\rm ms}/\Delta t}.
\label{G bound}
\ee
As was mentioned above, the best experimental bound on ${\lambda^2}/{M_\Delta^2}$ follows from the non-observation of oscillations of muonium atom \cite{Willmann:1998gd}. We infer this bound from the review paper \cite{Abada:2007ux} taking into account that the definition of coupling constants in \cite{Abada:2007ux} and in the present work differ by the factor $\sqrt{2}$:
\be\label{muonium constraint}
\frac{\lambda^2}{(M_\Delta/{\rm TeV})^2} < 0.2, ~~~ 90\%~{\rm CL}.
\ee
From inequalities (\ref{G bound}) and (\ref{muonium constraint}) it follows that even for $\Delta t=1$ ms there is a room for a singlet bilepton which affects the supernova physics and evades direct experimental constraints. If one takes $\lambda=0.1,$ then the bilepton mass should satisfy
\be
220 ~{\rm GeV} < M_\Delta \lesssim 3.2\cdot (300~{\rm ms}/\Delta t)^{-1/4} ~{\rm TeV}.
\ee



\section{Scalar triplet. See-saw type II.}
\subsection{Cross sections}
Let us now discuss LFV mediated by a scalar triplet $\bf \Delta$. We are especially interested in the case when such triplet is responsible for the generation of Majorana neutrino masses through the see-saw type II mechanism \cite{Magg:1980ut}-\cite{Mohapatra:1980yp}. In this mechanism scalar triplet additionally couples to the SM Higgs bosons, this coupling producing a vacuum expectation value for the neutral component of the triplet. The neutrino masses are proportional to this vev.

The see-saw type II Lagrangian contains two major ingredients,
a scalar-lepton interaction,
\be
{\cal L}_{ll\Delta}=\sum_{l,l'}\lambda_{l l'} \overline{ L_l^c} i\tau_2 \Delta L_{l'}+h.c.,
\label{llDelta}
\ee
and a scalar potential, which in its minimal form reads
\be
V= -M_H^2 H^\dagger H + f(H^\dagger H)^2+ M_\Delta^2 Tr (\Delta^\dagger\Delta) + \frac{1}{\sqrt{2}}(\tilde \mu H^T i\tau_2 \Delta^\dagger H+h.c.).
\label{Delta2}
\ee
Here
\be
\Delta\equiv {\bm \Delta \bm \tau}/\sqrt{2}=
\begin{pmatrix}
\Delta^+/\sqrt{2} & \Delta^{++}\\
\Delta^0 & -\Delta^+/\sqrt{2}
\end{pmatrix},
\ee
$L_l\equiv
\begin{pmatrix}
(\nu_{l})_L\\
l_{L}
\end{pmatrix}
$ is a doublet of left-handed leptons of flavor $l=e,\mu,\tau$,
$H$ is a Higgs doublet,
$\tilde \mu$ is a parameter with the dimension of mass.
Note that the definition of coupling constant $\tilde \mu$ varies in different works by a numerical coefficient. Our definition coincides e.g. with that in \cite{Akeroyd:2009nu}.

Note that due to the anticommutation of the fermion fields $3\times3$ matrix $\Lambda\equiv||\lambda_{ll'}||$ is symmetric,
\be\label{symmetry of lambda}
\Lambda^T=\Lambda.
\ee

The vev of the neutral component of the triplet reads
\be
\langle\Delta^0\rangle=\frac{\tilde\mu v^2}{2 \sqrt{2} M_\Delta^2},
\ee
where $ v\equiv \sqrt{2} \langle H^0\rangle = 246$ GeV. Due to the triplet vev neutrinos acquire the Majorana mass according to
\be
m = 2 \langle\Delta^0\rangle \Lambda,
\label{m-lambda relation}
\ee
where $m\equiv ||m_{l l'}||$ is a neutrino mass matrix in a flavor basis. One gets that in the see-saw type II model neutrino mass matrix $m$ is proportional to the coupling matrix $\Lambda.$ Therefore now $\Lambda$ can not be exactly diagonal as in the previous section, because $m$ has non-zero non-diagonal entries (we know this since neutrinos oscillate). However, current experimental data on neutrino masses and mixing are incomplete and  do not fix the ratio of diagonal and non-diagonal elements. In particular, it may appear that $m$ is approximately diagonal, and thus it is possible to achieve the suppression of the processes with $|\Delta L_l|=1$ compared to the processes with $|\Delta L_l|=2.$
This issue is considered in detail in the next subsection.

Lagrangian (\ref{llDelta}) gives rise to all four reactions (\ref{LFV processes relevant}), potentially relevant for supernova. Their cross sections in the center of mass reference frame read

\be\label{sigma 1}
\sigma(ee\rightarrow \mu\mu)=(|\lambda_{ee}|^2|\lambda_{\mu\mu}|^2/M_\Delta^4)(1-m_\mu^2/2E_e^2)\sqrt{1-m_\mu^2/E_e^2}~E_e^2/2\pi,
\ee
\be\label{sigma 2}
\sigma(e\nu_e\rightarrow \mu\nu_\mu)=(|\lambda_{ee}|^2|\lambda_{\mu\mu}|^2/M_\Delta^4)(1-m_\mu^2/4E_e^2)^2E_e^2/4\pi, ~~~~ E_e=E_{\nu_e},
\ee
\be\label{sigma 3}
\sigma(\nu_e\nu_e\rightarrow \nu_\mu\nu_\mu)=(|\lambda_{ee}|^2|\lambda_{\mu\mu}|^2/M_\Delta^4)E_{\nu_e}^2/2\pi,
\ee
\be\label{sigma 4}
\sigma(\nu_e\nu_e\rightarrow \nu_\tau\nu_\tau)=(|\lambda_{ee}|^2|\lambda_{\tau\tau}|^2/M_\Delta^4)E_{\nu_e}^2/2\pi,
\ee
where neutrinos and electrons are assumed to be extremely relativistic.

Density and chemical potential of electrons are higher than those for electron neutrinos; also an additional factor $1/2$ persists in eq. (\ref{sigma 2}). As a consequence, the first reaction in (\ref{LFV processes relevant}) dominates over the second and the third ones provided the c.o.m. energy is considerably greater than the  threshold $2m_\mu.$
The conditions on effective coupling constants, {\it either} of which ensures the appreciable rate of LFV in supernova core, now read
\be\label{condition 1}
\frac{|\lambda_{ee}\lambda_{\mu\mu}|}{M_\Delta^2} \gtrsim  10^{-3} \sqrt{300~{\rm ms}/\Delta t} ~\frac{1}{{\rm TeV}^2},
\ee
\be\label{condition 2}
\frac{|\lambda_{ee}\lambda_{\tau\tau}|}{M_\Delta^2} \gtrsim 2.4 \cdot 10^{-3} \sqrt{300~{\rm ms}/\Delta t} ~\frac{1}{{\rm TeV}^2}.
\ee
It is shown in the following subsection that $\lambda_{\mu\mu} \simeq \lambda_{\tau\tau}$ for the interesting set of neutrino parameters. Therefore  it is always easier to satisfy condition (\ref{condition 1}) than condition (\ref{condition 2}); for this reason we concentrate on condition (\ref{condition 1}) in what follows.

If one assumes $\lambda_{ee} \simeq \lambda_{\mu\mu} = 0.1$ and takes $\Delta t=300$ ms, then, analogously to the case considered in the previous section, supernova appears to be able to test $M_\Delta\lesssim 3.2$ TeV. To compare, the LHC can probe $M_\Delta\lesssim 1$ TeV \cite{Akeroyd:2007zv}.

\subsection{Neutrino mass matrix, rare decays and experimental constraints on TeV-scale see-saw type II}

LFV mediated by bileptons is constrained by non-observation of rare decays (\ref{rare decays}) and muonium conversion, see \cite{Abada:2007ux}\cite{Cuypers:1996ia}.  The widths of rare decays and the rate of conversion are proportional to $|\lambda_{l_1l_2}\lambda_{l_3l_4}/M_\Delta^2|^2,$ where $l_1-l_4$ stand for the corresponding leptons. Experimental bounds are presented in Table \ref{table rare processes}.


\begin{table}[t]
\vspace{0.2cm}
\begin{center}
 \begin{tabular}{|c|c|c|}
\hline
process & constraint on & bound, $\times (M_\Delta/{\rm TeV})^2$ \\
\hline
$\mu^-  \rightarrow  e^+e^-e^-$   &   $|\lambda_{e\mu}\lambda_{ee}|$    &   $ < 2.4 \cdot 10^{-5}$\\
\hline
$\tau^-  \rightarrow  e^+e^-e^-$   &   $|\lambda_{e \tau}\lambda_{ee}|$    &   $ < 2.6 \cdot 10^{-2}$\\
\hline
$\tau^-  \rightarrow  \mu^+\mu^-\mu^-$   &   $|\lambda_{\mu\tau}\lambda_{\mu\mu}|$    &   $ < 2.4 \cdot 10^{-2}$\\
\hline
$\tau^-  \rightarrow  \mu^+e^-e^-$   &   $|\lambda_{\mu\tau}\lambda_{ee}|$    &   $ < 1.9 \cdot 10^{-2}$\\
\hline
$\tau^-  \rightarrow  e^+\mu^-\mu^-$   &   $|\lambda_{e\tau}\lambda_{\mu\mu}|$    &   $ < 2.0 \cdot 10^{-2}$\\
\hline
$\mu  \rightarrow  e\gamma$   &   $|\lambda_{e\mu}^*\lambda_{ee}+\lambda_{\mu\mu}^*\lambda_{e\mu}+\lambda_{\tau\mu}^*\lambda_{e\tau}|$    &   $ < 9.4 \cdot 10^{-3}$\\
\hline
$\mu^-e^+  \leftrightarrow  \mu^+e^-$   &   $|\lambda_{\mu\mu}\lambda_{ee}|$    &   $ < 0.2$\\
\hline
\end{tabular}
\end{center}
\caption{
Constraints from low energy experiments. We extract numerical values from the review paper \cite{Abada:2007ux} taking into account that the definitions of coupling constants in \cite{Abada:2007ux} and in the present work differ by the factor $\sqrt{2}.$
\label{table rare processes}
			}
\end{table}
Evidently the most stringent constraint stems from $\mu  \rightarrow  eee$ decay. In the present subsection we first concentrate on this constraint and address the question what mass-mixing pattern fits it and simultaneously allows sizable LFV in supernova. Then we show that if this constraint does not contradict the condition (\ref{condition 1}), then other constraints also do not contradict it.

The $\mu  \rightarrow  eee$ constraint does not contradict the condition (\ref{condition 1}) if
\be\label{diagonal-nondiagonal ratio}
\left|\lambda_{\mu\mu}/\lambda_{e\mu}\right|=|m_{\mu\mu}/m_{e\mu}|\gtrsim 40 \sqrt{300~{\rm ms}/\Delta t} .
\ee

The neutrino mass matrix in the flavor basis, $m,$ is obtained from the diagonal mass matrix through the transformation \cite{Schechter:1980gr}
\be\label{mass matrix}
m= U^* \cdot {\rm diag}(m_1,m_2,m_3) \cdot U^\dagger,
\ee
where $U\equiv||U_{li}||$ ($l=e,\mu,\tau,~i=1,2,3$) is a PMNS neutrino mixing matrix:
$$
 U =
\left( \begin{array}{ccc} 1 & 0 & 0 \\ 0 & c_{23}  & s_{23}
\\ 0 & -s_{23} & c_{23} \end{array} \right)
\left( \begin{array}{ccc} c_{13} & 0 & s_{13} e^{-i\delta}\\ 0 & 1 & 0 \\
-s_{13}e^{i\delta} & 0 & c_{13}
\end{array} \right)
\left( \begin{array}{ccc}   c_{12} & s_{12}  & 0
\\  -s_{12} & c_{12} & 0 \\ 0 & 0 & 1 \end{array} \right)\times
$$
\be\label{PMNS matrix}
\times{\rm diag}(e^{i\alpha_1/2}, e^{i\alpha_2/2},1),
\ee
$$
\begin{array}{ll} c_{ij} \equiv \cos\theta_{ij} \; , & s_{ij}
\equiv \sin\theta_{ij} \;\; .
\end{array}
$$
Note that eq.(\ref{mass matrix}) implies $m^T=m$ in accordance with eqs.(\ref{symmetry of lambda}),(\ref{m-lambda relation}). The explicit expressions for the entries of $m$ may be found in \cite{Akeroyd:2007zv,Perez:2008ha,Garayoa:2007fw,Kadastik:2007yd}; for the sake of completeness we provide this expressions in our notations.\footnote{It is worth noting that if the Majorana phase matrix would stay at the first place in the r.h.s. of eq.(\ref{PMNS matrix}), then the absolute values of entries of $m$ would not depend on Majorana phases. The correctness of the ordering of matrices in eq.(\ref{PMNS matrix}) is justified by the diagonalization of lepton mass matrices \cite{Schechter:1980gr}.}
\be\label{m entries}
\begin{array}{lcl}
m_{ee} & = & a ~c_{13}^2  + s_{13}^2 m_3  e^{2 i \delta } \\
m_{\mu\mu} & = &  m_1 e^{-i \alpha_1} (s_{12} c_{23} + s_{13} e^{-i \delta
} c_{12} s_{23})^2+m_2 e^{-i \alpha_2} (c_{12} c_{23}-s_{13} e^{-i \delta } s_{12} s_{23})^2+m_3 c_{13}^2 s_{23}^2\\
m_{\tau\tau} & = & m_1 e^{-i \alpha_1} (s_{12} s_{23}- s_{13} e^{-i \delta } c_{12} c_{23})^2+m_2 e^{-i \alpha_2} (c_{12} s_{23}+ s_{13} e^{-i \delta }  s_{12} c_{23})^2+m_3 c_{13}^2 c_{23}^2\\
m_{e\mu} & = & c_{13} [d  s_{12} c_{12} c_{23} +s_{13} e^{i \delta }  s_{23} (m_3-a e^{-2 i \delta })]\\
m_{e\tau} & = & c_{13}[-d  s_{12} c_{12} s_{23} +  s_{13} e^{i \delta } c_{23} (m_3-a e^{-2 i \delta }) ] \\
m_{\mu\tau} & = & s_{23} c_{23}(-b+c_{13}^2  m_3) - s_{13} d e^{-i \delta}  s_{12}c_{12}(c_{23}^2-s_{23}^2) + s_{13}^2 a e^{-2i\delta} s_{23} c_{23}  \\
\end{array}
\ee
Here we define the parameters with the dimension of mass
\be
\begin{array}{lcl}
a   &   \equiv   &  m_1 e^{-i \alpha_1} c_{12}^2+m_2 e^{-i \alpha_2} s_{12}^2     \\
b   &   \equiv   &  m_1 e^{-i \alpha_1} s_{12}^2 + m_2 e^{-i \alpha_2}  c_{12}^2  \\
d   &   \equiv   &  m_2e^{-i \alpha_2} -m_1 e^{-i \alpha_1}                        \\
\end{array}
\ee
If the masses  $m_1$ and $m_2$ are quasi-degenerate, $m_1\simeq m_2,$  and Majorana phases $\alpha_1$ and $\alpha_2$ are approximately equal, $\alpha_1\simeq\alpha_2$ (this case is of particular interest, as is shown below), then this parameters reduce to

\be
a     \simeq b \simeq    e^{-i \alpha_1}  m_1
\ee
\be
d     \simeq     e^{-i \alpha_1} (m_2-m_1)                         \\
\ee

There are nine parameters in the mass matrix: three neutrino masses $m_i,$ three mixing angles $\theta_{ij},$ the Dirac phase $\delta$ and two Majorana phases $\alpha_{1,2}.$ Experimentally measured (or constrained) quantities are \cite{GonzalezGarcia:2007ib}:
\begin{equation}
\begin{array}{cccc}
\theta_{12} \simeq 34^o, &  \theta_{23}\simeq 45^o ~{\rm or}~ \theta_{23}\simeq 135^o , &\theta_{13}<13^o,
\end{array}
\end{equation}
\be
\begin{array}{cc}
\Delta m_{21}^2 \simeq 0.8 \cdot 10^{-4} {\rm eV}^2, & \Delta m_{32}^2 \simeq \pm 2.4 \cdot 10^{-3} {\rm eV}^2.
\end{array}
\ee
In addition, cosmological considerations 
\cite{Malinovsky:2008zz} demonstrate that
\be
m_1,m_2,m_3 < 0.35 {\rm ~eV}.
\ee
Unknown quantities are phases $\delta$ and $\alpha_{1,2},$ the mixing angle $\theta_{13},$ the sign of $\Delta m_{32}^2,$ the quadrant of $\theta_{23},$ the absolute scale of neutrino masses. Depending on the sign of $\Delta m_{32}^2$ two patterns of neutrino masses are possible: normal mass hierarchy (NH), $m_1<m_2<m_3,$ and inverted mass hierarchy (IH), $m_3<m_1<m_2.$

Formulas (\ref{m entries}) allow to check whether condition (\ref{diagonal-nondiagonal ratio}) is valid for any specific choice of neutrino masses and mixing parameters. However, let us outline general trends of the dependence of $|m_{\mu\mu}/m_{e\mu}|$ on the neutrino masses, mixing angles and phases.

Generically
\be\label{m(mumu), order of magnitude}
|m_{\mu\mu}| \sim  {\rm max}\{m_2,m_3\}.
\ee
If one takes all three phases, $\alpha_1,\alpha_2,\delta,$ randomly from the interval $[0,2\pi],$ then in the majority of cases
\be\label{m(emu), order of magnitude}
|m_{e\mu}| \sim {\rm max}\{m_2,s_{13}m_3\},
\ee
which implies $|m_{\mu\mu}/m_{e\mu}|\lesssim 5.$\footnote{To get this bound one should take into account that $m_2>0.009$ eV and in case of normal hierarchy $m_3>0.05$ eV.} Hence, in order to fit inequalities (\ref{diagonal-nondiagonal ratio}) one should first of all impose some bounds on the phases. Note that if all three phases vanish (which correspond to the exact CP-conservation in the neutrino sector), $m_{e\mu}$ reduces to
\be\label{m(emu) CP conservation}
m_{e\mu}\simeq  s_{12} c_{12} c_{23}\Delta m_{21}^2/(m_1+m_2)+s_{13} s_{23} \Delta m_{32}^2/(m_2+m_3).
\ee
The first term here is small compared to $m_2$ if $m_1\simeq m_2.$ The second term is small compared to $m_3$ if either $m_2\simeq m_3$ or $s_{13}$ is small. Therefore we come to a conclusion that small phases, large masses (which implies degeneracy of masses) and small $\theta_{13}$ favour LFV in the supernova core. This is analogous to the conclusions of ref.\cite{Malinsky:2008qn}, in which collider and low-energy LFV signatures were considered.

\begin{figure}[t]
\centerline{
\includegraphics[scale=0.7]{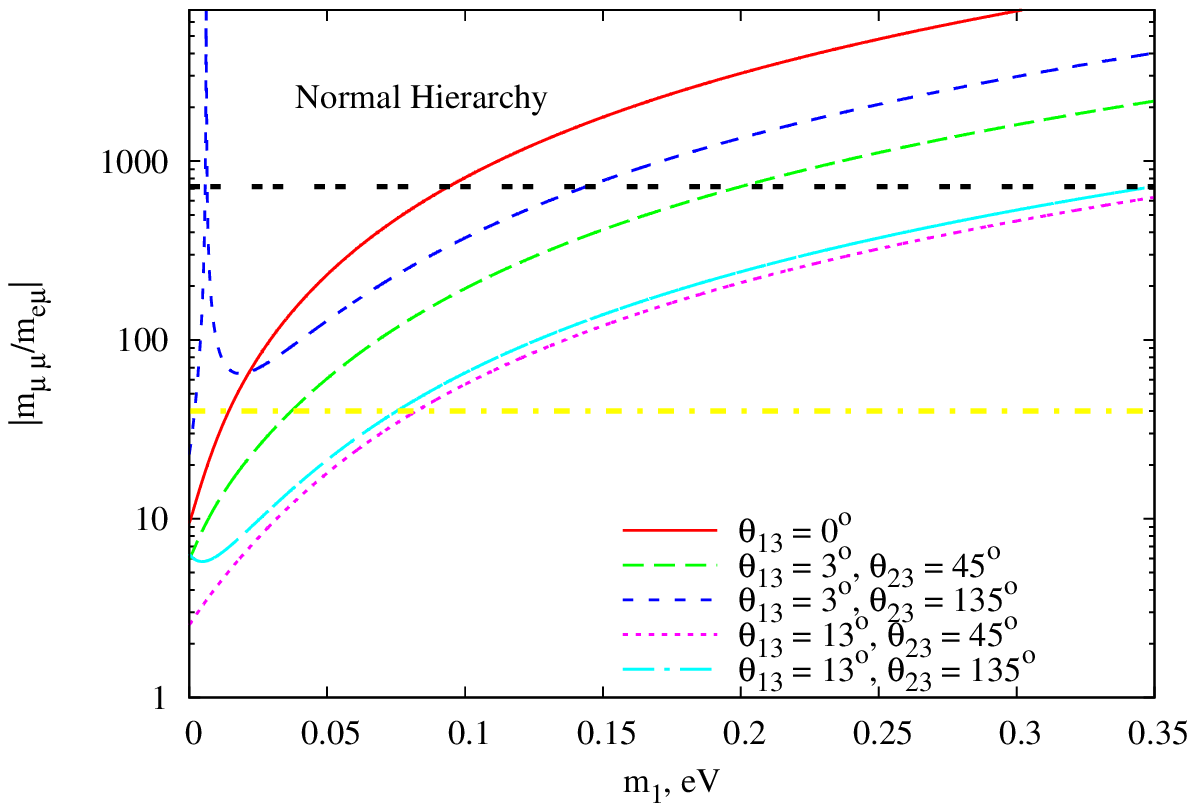}
\includegraphics[scale=0.7]{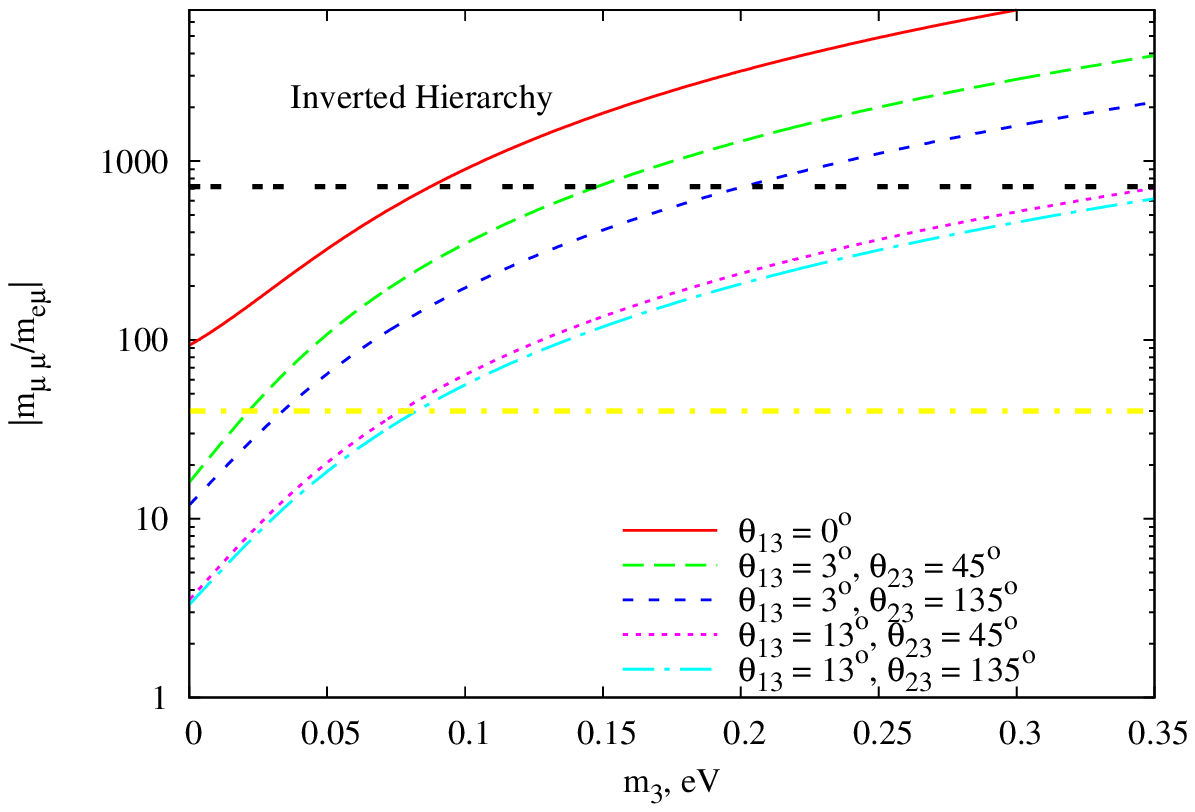}
}
\caption{\label{CP conservation} Ratio $|m_{\mu\mu}/m_{e\mu}|$
in case of exact CP-conservation ($\alpha_1=\alpha_2=\delta=0$).
 If $\theta_{13}=0$ then $|m_{\mu\mu}/m_{e\mu}|$ does not depend on the quadrant of  $\theta_{23}.$ Two thick horizontal lines correspond to $|m_{\mu\mu}/m_{e\mu}|=40 $ and $|m_{\mu\mu}/m_{e\mu}|=720$, which are minimal values sufficient for LFV in supernova core to proceed  at scales 300 ms and 1 ms accordingly.}
\end{figure}

\begin{figure}[t]
\centerline{
\includegraphics[scale=0.7]{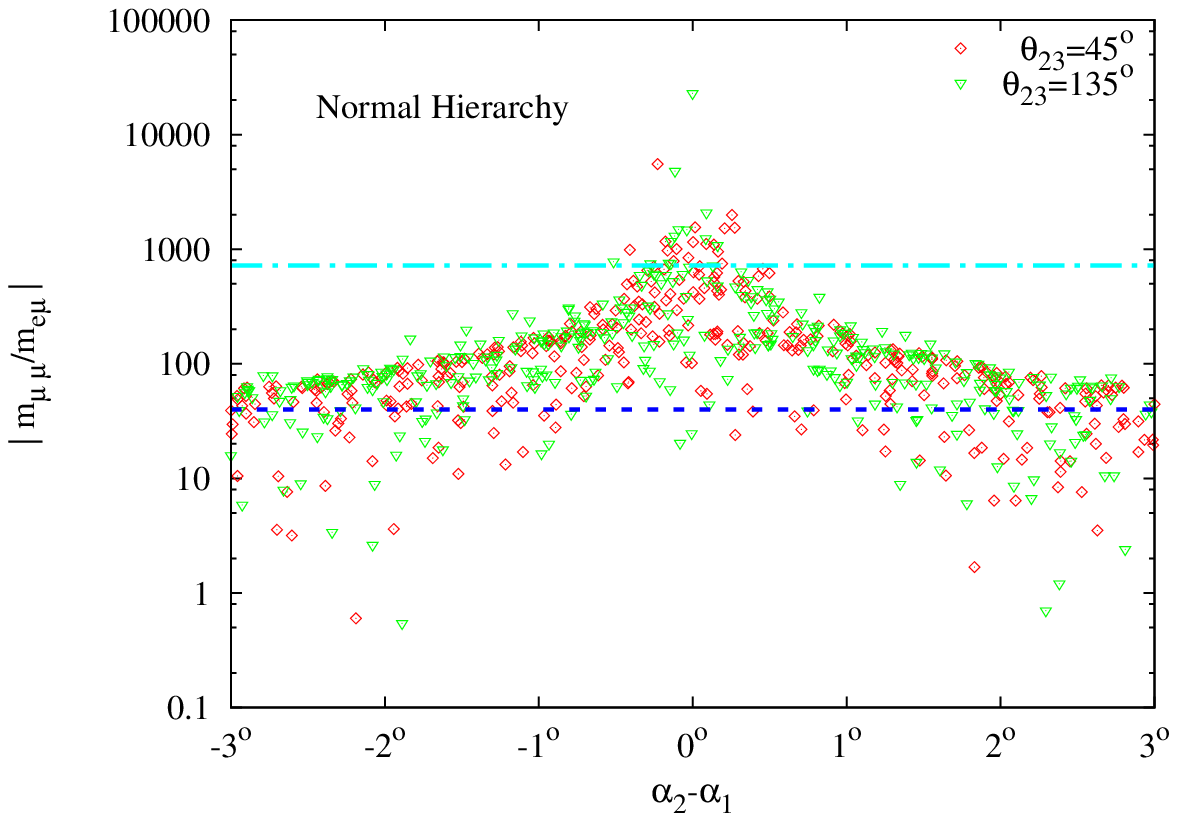}
\includegraphics[scale=0.7]{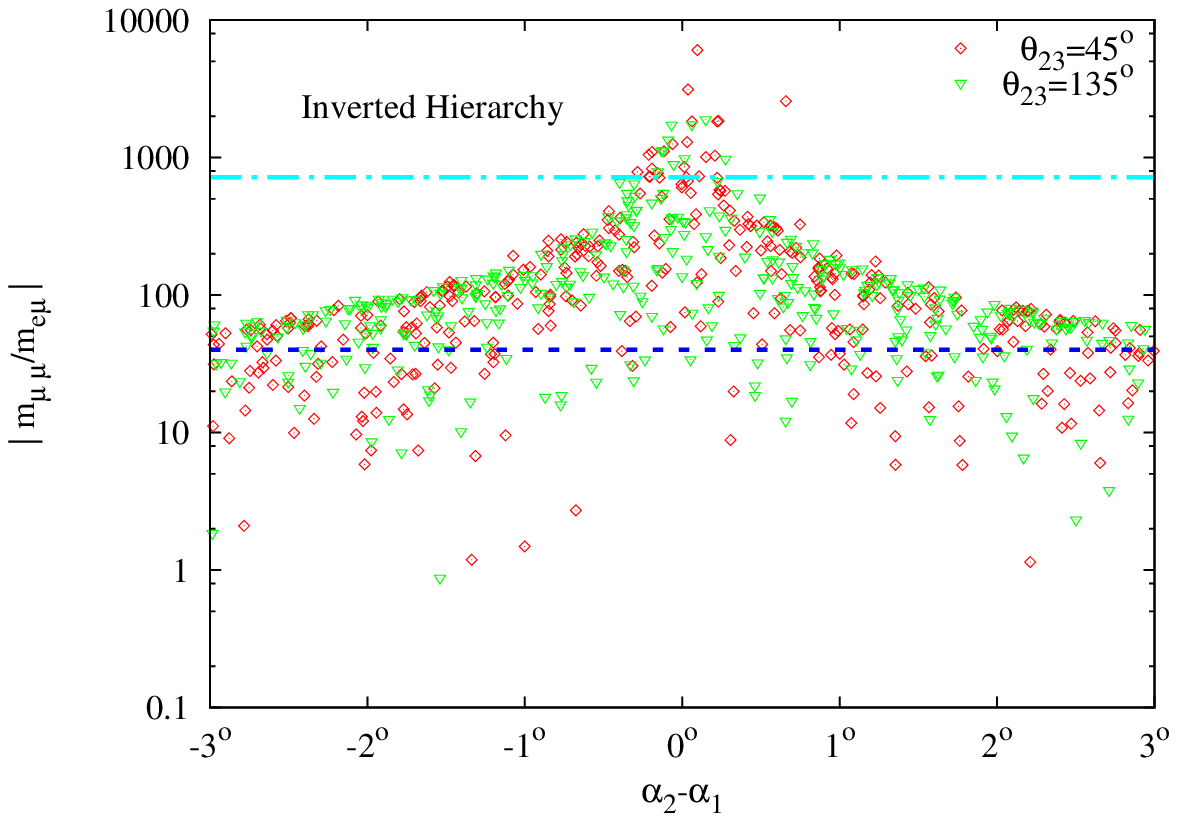}
}
\caption{\label{deviations} Scatter plots for the ratio $|m_{\mu\mu}/m_{e\mu}|.$  Each point corresponds to a random choice of neutrino parameters in the following ranges: $m_{\rm min}\in[0.05~{\rm eV},0.35~{\rm eV}], \theta_{13}\in[0^o,3^o],\delta\in[-90^o,90^o],\alpha_1\in[-2\delta-3^o,-2\delta+3^o],\alpha_2\in[-2\delta-3^o,-2\delta+3^o].$ If the hierarchy is normal, then $m_{\rm min}=m_1,$ otherwise $m_{\rm min}=m_3$}
\end{figure}

\begin{figure}[t]
\centerline{
\includegraphics[scale=0.7]{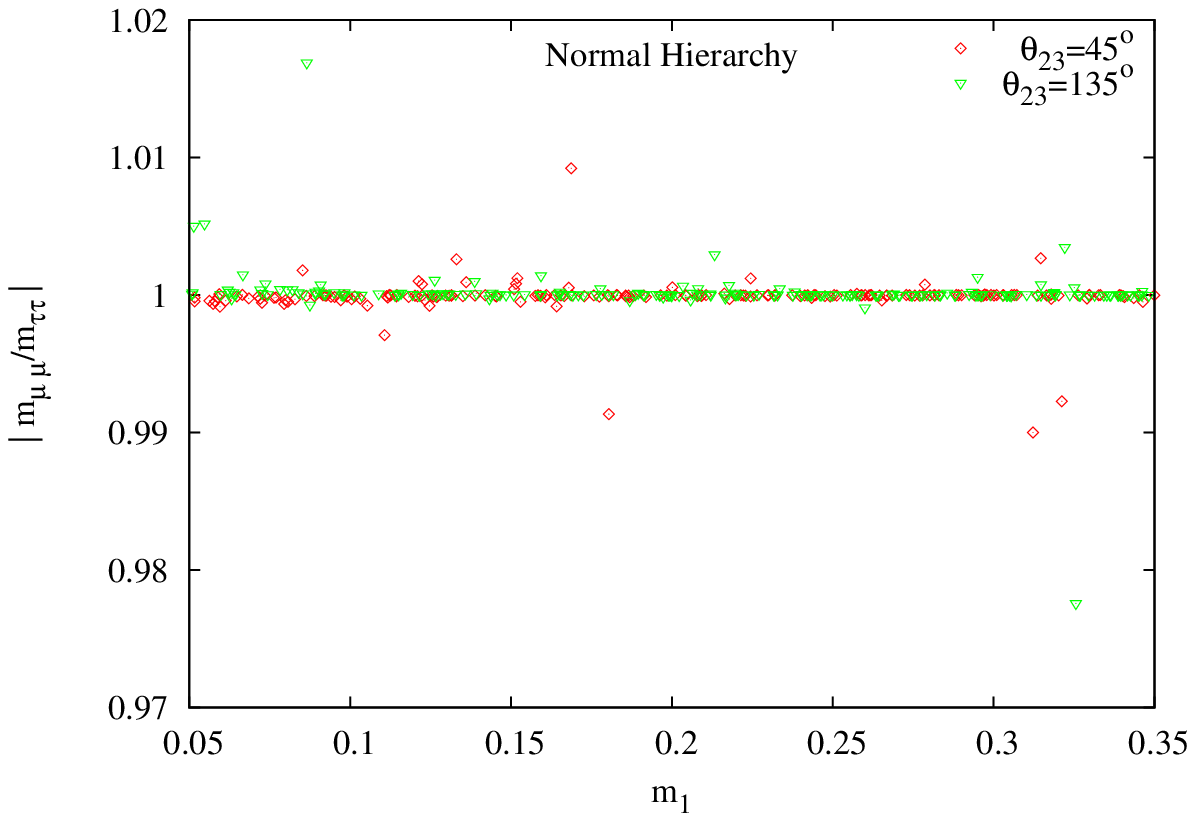}
\includegraphics[scale=0.7]{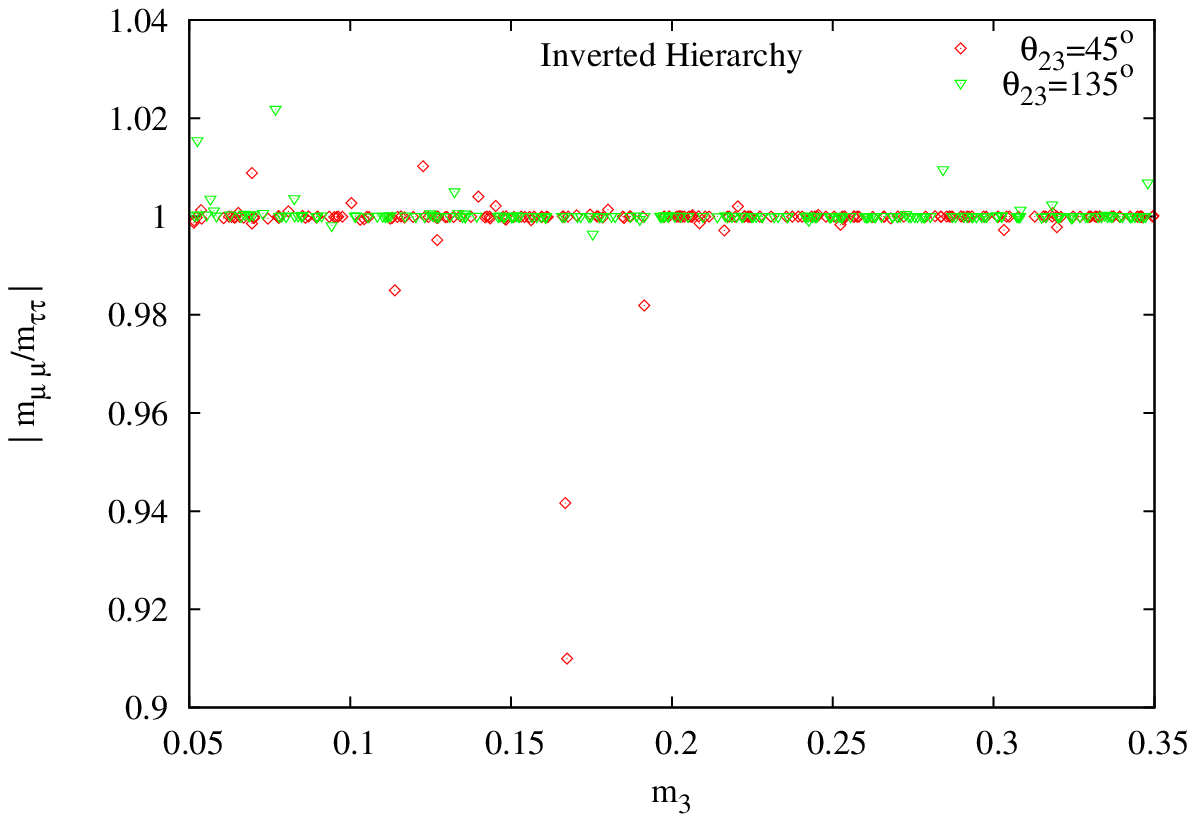}
}
\caption{\label{mumutautauratio} Scatter plots for the ratio $|m_{\mu\mu}/m_{\tau\tau}|$. Each point corresponds to a random choice of neutrino parameters in the following ranges: $\theta_{13}\in[0^0,3^0],\delta\in[-90^o,90^o],\alpha_1\in[-2\delta-3^0,-2\delta+3^0],\alpha_2\in[-2\delta-3^0,-2\delta+3^0].$ }
\end{figure}

\begin{figure}[p]
\centerline{
\includegraphics[scale=0.7]{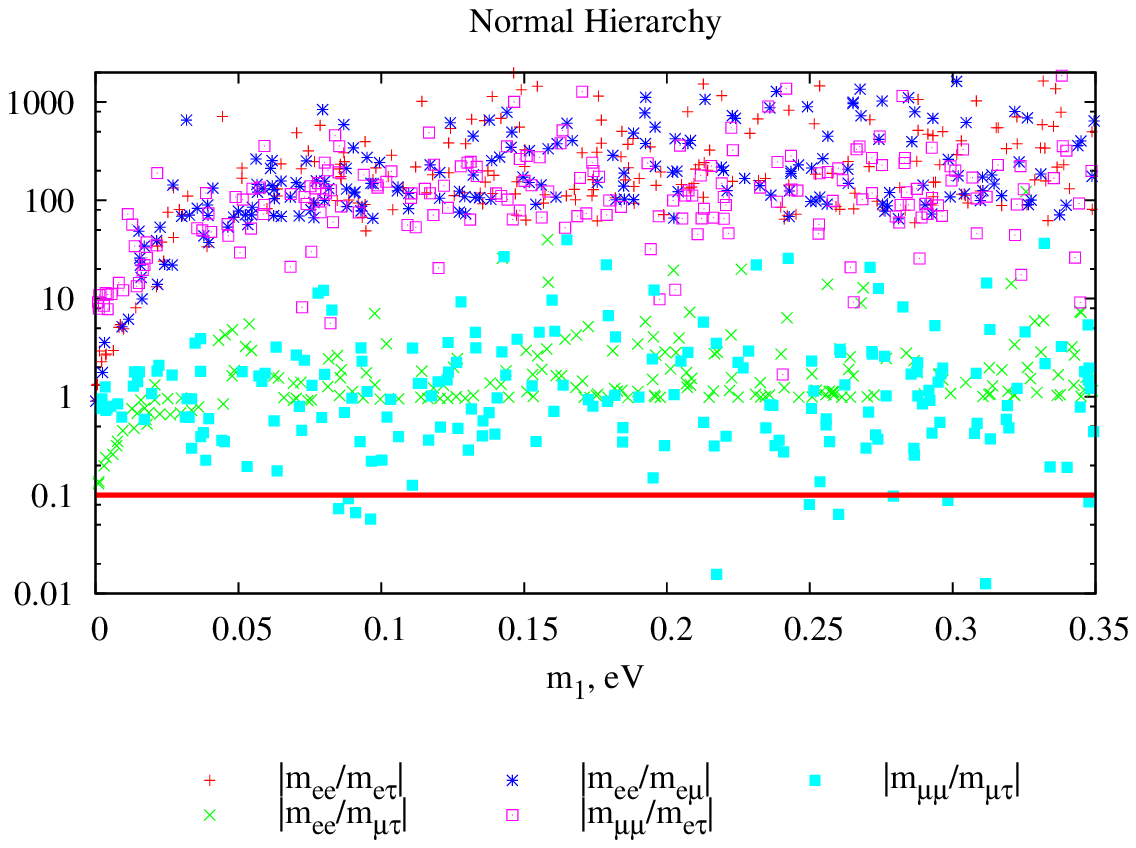}
\includegraphics[scale=0.7]{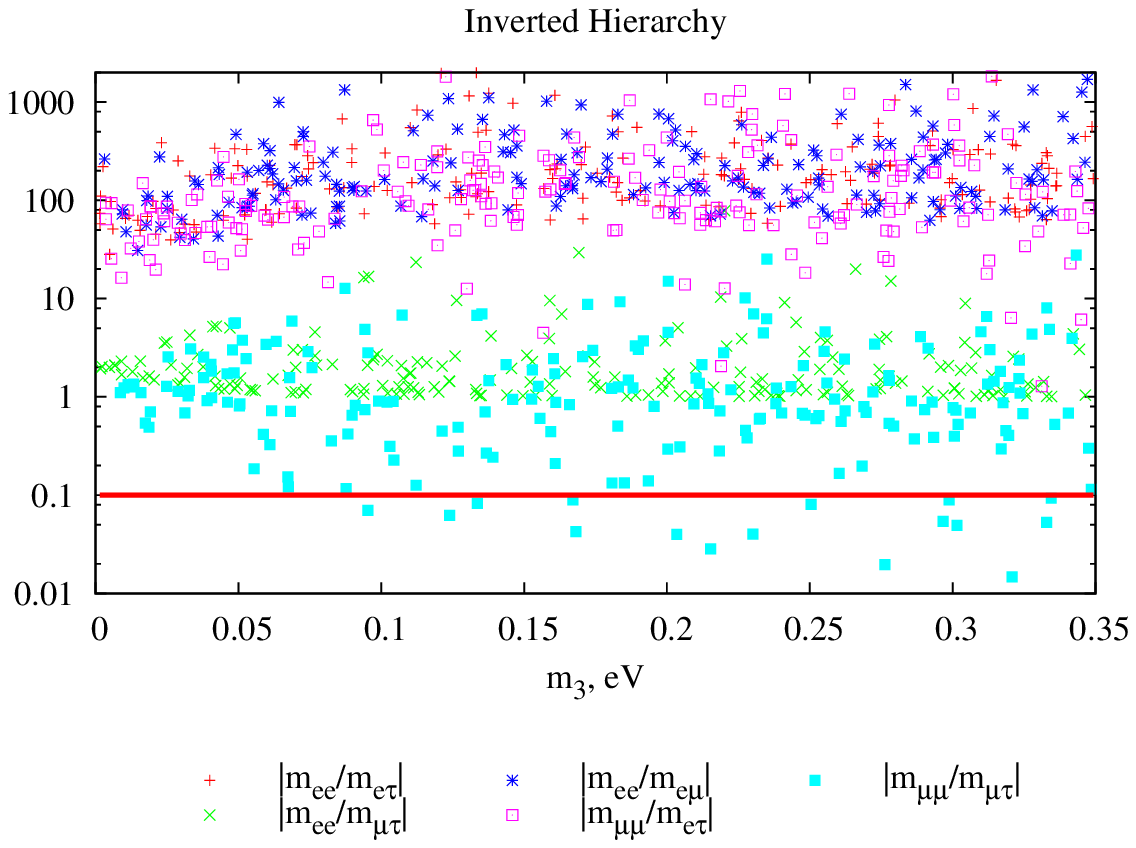}
}
\caption{\label{other} Scatter plots for the ratios $|m_{ee}/ m_{e\tau}|,|m_{ee}/ m_{\mu\tau}|,|m_{ee}/ m_{e\mu}|,|m_{\mu\mu}/ m_{e\tau}|,|m_{\mu\mu}/ m_{\mu\tau}|.$  Each point corresponds to a random choice of neutrino parameters in the following ranges:
$\theta_{13}\in[0^0,3^0],\delta\in[-90^o,90^o],\alpha_1\in[-2\delta-3^0,-2\delta+3^0],\alpha_2\in[-2\delta-3^0,-2\delta+3^0].$ The quadrant for $\theta_{23}$ is chosen randomly for each point. }
\end{figure}

\begin{figure}[p]
\centerline{
\includegraphics{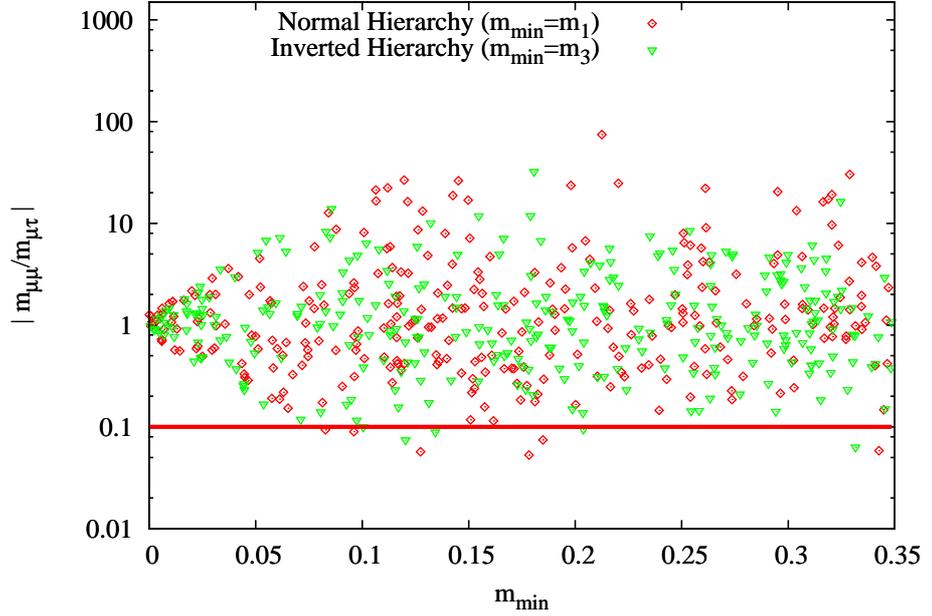}
}
\caption{\label{mumumutauratio} Scatter plot for the ratio $|m_{\mu\mu}/ m_{\mu\tau}|$ in case when $\alpha_1\simeq \alpha_2 \simeq -2 \delta$ and  $\delta \neq \pm\pi/2,~\alpha_i\neq\pm\pi.$  Each point corresponds to a random choice of neutrino parameters in the following ranges:
$\theta_{13}\in[0^0,3^0],\delta\in[-87^o,87^o],\alpha_1\in[-2\delta-3^0,-2\delta+3^0],\alpha_2\in[-2\delta-3^0,-2\delta+3^0].$ The quadrant for $\theta_{23}$ is chosen randomly for each point.}
\label{ee->mumu}
\end{figure}

We illustrate and refine this conclusions by means of plots depicted in Figs. \ref{CP conservation}-\ref{mumumutauratio}. On Fig. \ref{CP conservation} the ratio $|m_{\mu\mu}/m_{e\mu}|$ is plotted as a function of a minimal neutrino mass ($m_1$ for NH, $m_3$ for IH) for different values of $\theta_{13}$ and different choices of the quadrant of $\theta_{23}.$  All phases are taken to be zero,  $\alpha_1=\alpha_2=\delta=0.$ One can see that in case of  quasi-degeneracy of all three neutrino masses, $m_1\simeq m_2\simeq m_3 \gtrsim 0.1$ eV, condition (\ref{diagonal-nondiagonal ratio}) is fulfilled for any allowed value of $\theta_{13}.$ In case of a less stringent restriction, $m_1\simeq m_2 \gtrsim 0.05$ eV, condition (\ref{diagonal-nondiagonal ratio}) with $\Delta t=300$ ms is fulfilled for $\theta_{13}\lesssim 3^o.$ Note that two terms in $m_{e\mu}$ have different signs either in case of NH, $c_{23}<0$ or in case of IH, $c_{23}>0,$ see  eq.(\ref{m(emu) CP conservation}). This explains, particulary, the discontinuity at $m_1\simeq0.006$ eV in case of normal hierarchy,  $\theta_{13}=3^o,~\theta_{23}=135^o.$

To what extent is it possible to relax the requirement $\alpha_1=\alpha_2=\delta=0?$ To answer this question we first note the following facts.
\begin{itemize}
\item If
\be\label{alpha1=alpha2}
 \alpha_1=\alpha_2,
\ee
then the parameter $d$ reduces to $e^{-i\alpha_1}(m_2-m_1),$ as was pointed out above. If we for a moment put $\theta_{13}=0,$ then condition (\ref{alpha1=alpha2}) ensures that   $m_{e\mu}$ is generically small: $m_{e\mu}=c_{13} c_{12} s_{12} c_{23} e^{-i\alpha_1}(m_2-m_1) \Delta m_{21}^2/(m_1+m_2).$
\item If a more restrictive condition is superimposed,
\be\label{alpha=-2delta}
 \alpha_1=\alpha_2=-2\delta,
\ee
then $m_{e\mu}$ appears to be a sum of two generically small terms (with some relative phase factor) even for $\theta_{13}\neq0:$
\be
m_{e\mu}\simeq c_{12} s_{12} c_{23}e^{2i\delta}\Delta m_{21}^2/(m_1+m_2)+s_{13} e^{i\delta} s_{23} \Delta m_{32}^2/(m_2+m_3).
\ee
\end{itemize}

Condition (\ref{alpha=-2delta}) is more general than the requirement $\alpha_1=\alpha_2=\delta=0$, and embraces the latter as a specific case. We illustrate allowed deviations from this condition on Fig.\ref{deviations}. One can see that equalities (\ref{alpha=-2delta}) should be valid up to $\sim 3^0$ for  $\Delta t=300$ ms in case when $\theta_{13}\lesssim 3^0$ and $m_{\rm min} \gtrsim 0.05$ eV. One may further relax condition (\ref{alpha=-2delta}) imposing more restrictive requirements on the degeneracy of neutrino masses and smallness of $\theta_{13}.$ On the other hand the specific value of $\alpha_1 \simeq \alpha_2 \simeq -2\delta,$ the quadrant of $\theta_{23}$ and the hierarchy of masses generically do not affect the value of $|m_{\mu\mu}/m_{e\mu}|$ significantly.

Now it is easy to check that $\lambda_{\mu\mu}\simeq\lambda_{\tau\tau}$  in the interesting range of neutrino mass-mixing parameters -- a claim made in the end of the previous subsection. This statement is illustrated by Fig.\ref{mumutautauratio}.

Finally we should demonstrate that as far as  condition (\ref{diagonal-nondiagonal ratio}) is fulfilled, not only \mbox{$\mu\rightarrow eee$} constraint but all other constraints from the rare decays are satisfied. Evidently, in order for the constraints cited in Table 2 to be compatible with condition (\ref{condition 1}) one needs
\be
|m_{\mu\mu}/ m_{e\tau}|,|m_{\mu\mu}/ m_{\mu\tau}|,|m_{ee}/ m_{e\tau}|,|m_{ee}/ m_{\mu\tau}|,|m_{ee}/ m_{e\mu}|\gtrsim 0.1.
\ee

Scatter plots for this ratios are presented on Fig.\ref{other}. Evidently all the ratios under consideration generically satisfy the above constraint. However, the ratio $|m_{\mu\mu}/ m_{\mu\tau}|$ appear to be smaller than 0.1 for a few choices of parameters.  Let us consider this ratio separately in more detail. Taking into account that $c_{23}^2=s_{23}^2$ one gets in the limit $\theta_{13}=0, m_1\simeq m_2\simeq m_3\gtrsim 0.1, ~\alpha_1=\alpha_2$
\be
\frac{|m_{\mu\mu}|}{| m_{\mu\tau}|} \simeq \left|\frac{m_3 +m_2 e^{i\alpha_2}}{m_3 - m_2 e^{i\alpha_2}}\right|.
\ee
If $\alpha_1=\alpha_2=\pi,$ this expression is small, $|m_{\mu\mu}/ m_{\mu\tau}|\simeq|\Delta m_{32}^2/4 m_3^2|\lesssim 0.006.$ Therefore one should demand $\alpha_1,\alpha_2\neq\pi$ in addition to eq. (\ref{alpha=-2delta})  in order not to fall in contradiction with $\tau\rightarrow\mu\mu\mu$ experimental constraints. Numerical analysis (see Fig.\ref{mumumutauratio}) shows that to evade the experimental constraint it is sufficient to have $|\alpha_i-\pi|$ greater than few degrees. The above considerations demonstrate that the $\tau\rightarrow\mu ee$ rare decay is the most useful one among all $\tau\rightarrow lll$ decays for the purpose of constraining the impact of LFV on the supernova physics. The importance of the $\tau\rightarrow\mu ee$ measurement for probing the see-saw type II model was emphasized in \cite{Fukuyama:2009xk}.

%

\section{Possible consequences of LFV  on supernova physics}
\label{LFVinSN}

\subsection{Standard collapse}
\label{Standard collapse}

Before discussing the influence of LFV on supernova physics, we first give a sketchy qualitative
picture using a standard approach (cf., e.g. \cite{Lattimer:1988}).

There are several stages in standard core-collapse scenario where different neutrino
processes dominate: 1) neutronization in the transparent core at infall before bounce;
2)~bounce, followed by shock-breakout with the most powerful neutrino burst;
3) accretion onto growing proto-neutron star with a stalled shock and possible shock
revival;
4) Kelvin-Helmholtz stage of cooling of a hot proto-neutron star.
The latter stage is characterized by a slow contraction of the hot core on a timescale
of many seconds with the release of gravitational energy through neutrino emission.
Contraction as a source of stellar luminosity was suggested by Kelvin and Helmholtz before the
discovery of nuclear reactions.
For the case of neutron cores inside stars this mechanism of stellar energy was considered
by Landau~\cite{Landau:1938}.

For a polytropic equation of state
\begin{equation}
P = K\rho^\gamma,
\label{polytr}
\end{equation}
the fundamental critical value of $\gamma$ is 4/3 in Newtonian theory. If the average
$\gamma < 4/3$ over a stellar core, then this core is dynamically unstable.
The only mass that may have a hydrostatic equilibrium for $\gamma=4/3$
in Newtonian gravity is the Chandrasekhar mass
\begin{equation}
M_{\rm Ch}=2.018(4\pi)\left(\frac{K}{\pi G}\right)^{3/2}.
\label{Chandr}
\end{equation}
This relation is derived in all classical textbooks on stellar structure.
Physically, it is explained by a chain of simple arguments.
The pressure $P$ in general is the flux of momentum $p$,
$P \sim pnv$, where $n$ is the number density and $v$ is the velocity of particles.
For extremely relativistic (ER) particles  $P \sim pnc$, and for degenerate electrons
$p \sim p_{\rm Fermi} \sim 1/\mbox{distance} \sim n^{1/3}$.
By definition of $Y_e$ -- the electron number density per baryon, $n=\rho Y_e/m_n$,
where $m_n$ is nucleon mass.
This implies
$P \propto n^{4/3} = (\rho Y_e/m_n)^{4/3}$ -- which means $\gamma=4/3$ in
relation (\ref{polytr}).
(The definition of $\gamma$ is $\partial \log P/ \partial \log \rho|_S$.)

We see that $K\propto (Y_e/m_n)^{4/3}$ in (\ref{polytr}).
On the other hand, gravity requires the equilibrium pressure
in the star center to be
$ \mbox{force}/\mbox{area} \propto G_N M^2/R^4
\propto G_N M^{2/3}\rho^{4/3} $.
When equated with $P = K\rho^{4/3}$ this fixes $M_{\rm Ch}$  in (\ref{Chandr}).
Since $K\propto (Y_e/m_n)^{4/3}$ one finds $M_{\rm Ch}\propto (Y_e/m_n)^2$.
When the coefficients are substituted accurately, one gets
$$
M_{\rm Ch} = 5.8 Y_e^2 M_\odot .
$$
Moreover, since, in natural units $G_N \sim m_{Pl}^{- 2}$, we have
an elegant expression
$$
 M_{\rm Ch}\sim  \frac{m_{Pl}^3 Y_e^2}{ m_n^2} ,
$$
where $m_{Pl}$ is Planck mass.

In general, instead of $Y_e$ there should be a sum over all particles producing pressure (e.g. $e$ and $\nu$).
And instead of the nucleon mass $m_n$ there may be a mass of another
particle if the latter dominates in producing gravity.
E.g. if the Dark Matter would be dominated by a light fermion with mass $m_1$
the limiting mass  $\sim  m_{Pl}^3 / m_1^2$ would be on a scale
of supercluster of galaxies if $m_1 \sim 1$~eV \cite{BKNov:1980}.

The contribution of baryons to the pressure is not
significant until $\rho > \rho_n \approx 3 \times 10^{14}$
(nuclear matter density).
The equation of state during collapse is
almost polytropic with $\gamma$ slightly lower 4/3 due to small corrections from
nuclei.
The value of $K$, which depends on $Y_L=Y_e +Y_\nu $ and entropy $S$, is
nearly constant since neutrinos are trapped.

Of course, at the bounce and later, in the center of the core the
pressure is dominated by nuclear forces.
However, the initial collapse begins when the pressure is
dominated by electrons.
The onset of the collapse in standard picture is determined either by the drop
of adiabatic exponent $\gamma$ below 4/3 due to iron dissociation or due to
neutronization (or both, and in both cases the collapse begins due to
the lowering pressure when the star core contracts).

The dynamics of collapse in this case can be described by
self-similar solutions \cite{SelfSim} which
show that a collapsing core
 separates into a homologous ($v\propto r$ ) central part,
and an outer, supersonically infalling part.
When the collapse is halted and bounces, a shock wave is produced at
the sonic point.
There is a self-similar solution with the shock as well \cite{AntonovaKazhdan:2000}.

A standard estimate for the free-fall collapse timescale \cite{Lattimer:1988} gives
\begin{equation}
t_f = \left(\frac{3}{8\pi G_N \rho}\right)^{1/2} \simeq 1.2 \rho_{12}^{-1/2}
\mbox{ms} ,
\label{tff}
\end{equation}
where $\rho_{12}$ is the average density in units of $10^{12}$ \mbox{g/cm$^3$}.
Below we use notation $\rho_k$ for  density in units of $10^{k}$ \mbox{g/cm$^3$}.

This estimate is not very realistic because it assumes zero pressure.
In reality the force produced by the pressure of ER degenerate leptons ($\sim n^{4/3} R^2$) and  the gravity force ($\sim G M^2/ R^{2} $) scale according to the
same law ($\propto R^{-2}$ ). Therefore the contribution of the  pressure force remains finite and non-negligible compared to the gravity force
 during the entire collapse.
A better estimate for the collapse timescale $t_c$ which agrees with numerical simulations
can be derived from self-similar models \cite{SelfSim,AntonovaKazhdan:2000}:
\begin{equation}
t_c \simeq 6 t_f .
\label{tff6}
\end{equation}

The mass of the inner core turns out to be about
10\% greater than the Chandrasekhar mass for the same
value of $K$, and is thus proportional to the lepton fraction
$Y_L^2$.

Another important timescale characterizes neutrino diffusion.
Only when the diffusion is less rapid than
the collapse will the lepton fraction be frozen for the
duration of the collapse. The standard expression for the
diffusion timescale is
\begin{equation}
t_d = \frac{3R^2}{c \ell_\nu}\simeq \frac{M^{2/3}\rho^{1/3}\sigma}{c m_n}.
\label{tdiff}
\end{equation}
If one takes the values corresponding to the inner core already after bounce, $M\simeq 0.6 M_\odot,~\rho \sim \rho_{14},~\sigma\sim 10^{-39}$~cm$^2$, one gets
\be
t_d \sim 10 ~ \mbox{s}.
\ee
Kelvin-Helmholtz time is a bit larger than the diffusion timescale because $t_d$
must be multiplied by a factor reflecting the total  energy storage over
the energy in neutrinos.

Neutrino signal basically consists of three consecutive periods. The first one is the neutronization burst -- a short ($\sim 10$ ms) powerful outburst of neutrinos which emerges when the shock wave reaches the neutrino sphere. Neutronization burst takes place few ms after bounce, and contains almost exclusively electron neutrinos, which are created during neutronization of stellar matter. The second period coincides with the existence of stalled shock and lasts $\sim 1$ s. The third period corresponds to the cooling of the proto-neutron star; it is characterized by an exponential decrease of neutrino luminosity and energy and lasts $\sim10$ s. At the second and the third stages neutrinos and antineutrinos of all flavors are in approximate equilibrium at the neutrino sphere (not inside the core!) and therefore are emitted approximately in equal quantities.

Neutrino oscillations take place {\it above} the neutrino sphere. In the layer $\sim 100$ km above the neutrino sphere collective oscillations occur (see e.g. \cite{Raffelt:2007xt}). In the outer stellar envelope more familiar MSW resonant flavor conversion take place (see e.g. \cite{Dighe:1999bi}). Oscillations modify the flavor content of the neutrino signal. In particular, they may convert a fraction of electron neutrinos of neutronization burst into non-electron neutrinos.

\subsection{LFV and collapse dynamics}

The rate of LFV reactions
(\ref{LFV processes relevant}) during
the \textit{infall} phase is negligible, with possible exception of the last few milliseconds. There are two reasons for this.
First,  the cross-sections
 (\ref{sigma 1})--(\ref{sigma 4}) are small being proportional to the
typical energies of $e$ and $\nu_e$ (i.e., to chemical potentials of $e$ and $\nu_e$).
Second, the rates $ n_{e,\nu_e} \langle \sigma v \rangle $ are small
when the concentrations $ n_e$ and $n_{\nu_e}$ are low.
Moreover, the first two reactions in  (\ref{LFV processes relevant}) have thresholds
proportional to muon mass $m_\mu$.
The reactions (\ref{LFV processes relevant}) may become important
the few ms before bounce
when the densities in the center approach the nuclear matter density.

Let us estimate, for example, when $ee \to \mu\mu$ process becomes important.
The reaction is possible when $E_e > m_\mu$.
At low density an electron may obtain such high energy only due to an exponential factor like
$\exp (-m_\mu/T)$, 
however $T$ is always  appreciably lower than  $m_\mu.$ At higher densities electrons become degenerate, and we should estimate their chemical potential, taking into account that electrons are extremely relativistic:
\begin{equation}
 \mu_e \approx 11(\rho_{10}Y_e)^{1/3} \; \mbox{MeV}.
\label{mue}
\end{equation}
We see that  $\mu_e$ surpasses the muon mass at $\rho \approx 3\times 10^{13}$ \mbox{g/cm$^3$} for $Y_e=0.3$  and the process
$ee \to \mu\mu$ sets in.
However, if $\mu_e$ is only slightly above  $m_\mu$, only a small fraction of
total electron density $n_e=\rho Y_e/m_n$ contributes to
$ n_e \langle \sigma v \rangle$ of this reaction: only electrons near and above the
Fermi surface take part in the reaction.

Nevertheless, the process of transformation $ee \to \mu\mu$ starts decreasing the electron
pressure already at $\rho_{13}  \sim 1$, and becomes very important at
$\rho_{13} \sim 10$, when the
chemical  potential of electrons is twice the muon mass and the bulk
of electrons contributes to the reaction.
At this stage we may safely use an estimate for the reaction rate
\begin{equation}
 \frac{1}{t} \sim n_e \langle \sigma v \rangle \approx 300 \rho_{14} Y_e \; \mbox{s}^{-1} ,
\end{equation}
where we take $E_e=2 m_\mu,$ $\lambda=0.1,$ and $M_\Delta =1$~TeV (this gives $\sigma \approx 2\times 10^{-46} \mbox{cm}^{2}$).
Thus the timescale for the process $ee \to \mu\mu$ is about 10 ms a few ms before
the bounce.
Our computations with the {\sc boom} code show  that the density  $\rho \sim 10^{13}$ \mbox{g/cm$^3$}
is achieved in the
center about 5 to 7 ms before the bounce in good agreement with estimate (\ref{tff6}),
and there is 1 to 2 ms from $\rho \sim 10^{14}$
\mbox{g/cm$^3$} until the bounce.




The disappearance of electrons in $ee \to \mu\mu$ reaction
starts actively contributing to the decrease of electron pressure -- hence to
the dynamics of the collapse.
Though the effect is rather small (in a couple of ms we lose only 10 to 20\% of electrons)
still all small deviations of $\gamma$ from 4/3 may play a role.
Moreover, the process of conversion will continue on later stages.
It will lead to the decrease of electron (and electron neutrino) chemical potential and to the growth of entropy and temperature, since electrons are taken away well below
the Fermi surface (cf. \cite{BisnKog:1970,Nakazawa:1973}).

It is interesting to estimate when the reverse
process $\mu\mu \to ee$ becomes significant in order to speak about
$\gamma$ in strict sense.
See Appendix on equilibrium number of muon pairs in standard physics scenario
when the muons would form a non-degenerate NR gas.

The two-way LFV reaction
$ee \rightleftharpoons \mu\mu$ leads eventually to the establishment of
equality $\mu_\mu = \mu_e$ for the chemical potentials. Muons become moderately degenerate since $\mu_\mu-m_\mu$ becomes somewhat larger than $T.$
When the equilibrium is already established, the muon gas has  $\gamma \simeq 5/3$ ($ P \sim pn_\mu v$ and $v=p/m_\mu$, hence
$P \sim p^2 n_\mu/m_\mu \propto n_\mu^{5/3}/m_\mu$), which is harder than for ER leptons, although softer than for
nuclear matter of high density.
However, in the beginning
of the transformation  $ee \to \mu\mu,$ when the reverse process is not yet effective,
the pressure of the muon gas is much smaller than that of the electron gas due to the large mass (and therefore low momenta) of muons. Thus in the beginning of the $ee \to \mu\mu$ conversion, which may start several ms before the standard bounce, a decrease of $P_e$ can not be compensated by the increase of $P_\mu,$ and we
have an effect of reduction of total pressure similar  to neutronization.
%
%
This may lead to a faster collapse, hence to a stronger bounce.

There is an opposite conclusion in \cite{Fuller:1987}.
They claim that the reduction of $Y_e$ (accompanied by the reduction of $P_e$) leads to a weaker bounce.

A \textit{one-zone} collapse model was used in \cite{Fuller:1987}. It may be good for
a uniform homologously contracting core where $\gamma$ is close to 4/3.
Even when electrons are transformed into trapped electron neutrinos the gamma for lepton gas is
close to 4/3 (all leptons are extremely relativistic). In this framework they give a standard estimate of the mass of the homologous core (HC), which is just approximated by a Chandrasekhar mass $M_{\rm Ch}:$ $M_{\rm HC} \simeq M_{\rm Ch} \simeq 5.8 Y_e^2 M_\odot$ (a bit more accurate estimate following  \cite{SelfSim,Lattimer:1988} is $M_{\rm HC} \simeq 1.1 M_{\rm Ch} $).




However, if the core started collapsing and then $Y_e$  drops fast enough and non-uniformly (e.g., faster in the center, as in our case), then the picture may be more complicated.

Effective $\gamma$ lower than 4/3 means faster collapse, hence a partial
loss of homology.
Quantitatively, the effect will be small if the process $ee \to \mu\mu$
is very slow. Anyway, here we see the qualitative difference with the standard case
which may lead to the deformation of the homologous flow at infall
and even to the formation of a second strong shock (because
the innermost parts of the flow, those that are in the deepest
potential well, have to accelerate due to the decrease of pressure).
We postpone the discussion of quantitative results for future research
keeping in mind possible new dynamical phenomena.

Another place where the reduction of  $\gamma $ may be interesting
is the shock after
the bounce. There is a stage of more or less steady  accretion through
a stalled shock.
When the pressure in the shock is dominated by leptons, and not by
nuclear matter, the reduction
of $\gamma $ means higher compression (the density jump in any strong
shock is $(\gamma+1)/(\gamma-1)$. However, when the free path corresponding to LFV processes (\ref{LFV processes relevant}) is much larger than a hundred of km this effect will not be noticeable. 

Above we mostly concentrated on the LFV conversion between electrons and muons.
A related questions should be answered if electron neutrinos equilibrate with muon and/or tau neutrinos. How does this affect supernova dynamics? Does this help the supernova to explode, or prevent the successful explosion?

A comprehensive analysis of similar problems may be found in paper~\cite{Rampp:2002kn}. A lepton number violating reaction  $\nu_e N\rightarrow\bar\nu_e N$ is considered there.
As in \cite{Fuller:1988} it is found that the decrease of lepton degeneracy leads to
a weaker shock wave, which prevents the prompt explosion (which is anyway disfavored by
numerical simulations). Another consequence is the increase of entropy and temperature of
the core. In principle this could affect the delayed explosion scenario.
However, the authors of \cite{Rampp:2002kn} state
on the basis of numerical simulation that the effect of LFV on the shock wave revival is
negligible. A difference between \cite{Rampp:2002kn} and present work  should be pointed out: $\nu_e\leftrightarrow\bar\nu_e$ transition results in a zero chemical potential of electron neutrinos with no compensation in other neutrino flavors, while LFV transitions considered here produce degenerate seas of muon and/or tau neutrinos at cost of reducing chemical potentials and concentrations of electrons and electron neutrinos. For this reasons the results of \cite{Rampp:2002kn} can not be directly applied for LFV processes.

\subsection{LFV and supernova neutrino signal}

First let us consider whether modifications of the neutronization burst are possible. Neutronization burst consists of neutrinos sweepped out by the shock wave from the {\it outer} core. On the other hand, as discussed above, the LFV reactions (\ref{LFV processes relevant}) may proceed effectively only in the {\it inner} core where the density and lepton chemical potentials are sufficiently high. Therefore it does not seem plausible that the flavor content of the standard neutronization burst may be altered. However, as pointed out in the previous subsection, LFV may modify the shock wave itself, which may cause the modification of the form, duration and total energy of the neutronization burst. If additional shock waves are produced due to inhomologous collapse, they result in additional neutronization bursts, in striking contrast to the standard picture.

Now let us turn to the neutrino emission after the neutronization burst. Its spectrum and flavor content is determined solely by the equilibrium conditions on the neutrino sphere. How can the redistribution of chemical potentials between different flavors inside the inner core affect such emission? To answer this question numerical simulations are necessary; here we propose only one possible signature. LFV leads to lowering of the average neutrino energy; also a fraction of electron neutrinos (which take part both in neutral current and in charge current interactions) is converted to non-electron neutrinos (which take part only in neutral current interactions). Both this circumstances tend to increase mean free path of neutrinos and to accelerate the cooling of the core. On the other hand, a decrease of neutrino degeneracy tends to decrease mean free path  and to decelerate the cooling. An interplay of this opposite trends may modify the total duration of the neutrino signal. This signature may be easily revealed by modern neutrino detectors, which should detect thousands of neutrino events in case of galactic supernova explosion.




\section{Summary}

We have shown that the lepton flavor violation, generated by the exchange of TeV-scale scalar bileptons, may drastically modify conditions inside a core-collapse supernova. An important example of such bileptons is a scalar triplet responsible for the neutrino mass generation in the see-saw type II scenario.  In this scenario the matrix of scalar-lepton couplings is proportional to the neutrino mass matrix.  The latter is currently partially fixed by neutrino oscillations and cosmological data. Also the effective four-fermion couplings, generated by the triplet, are restricted by low-energy experiments. We analyze the body of data and  demonstrate that there exists an allowed region of the see-saw II parameter space in which LFV is strong enough to lead to the thermal equilibrium between electron and non-electron species inside the collapsing core. Roughly speaking, LFV affects the supernova physics if the matrix of scalar-lepton couplings is approximately proportional to the unit matrix, $\lambda_{ll'} \simeq \lambda \delta_{ll'},$ and the effective coupling $\lambda^2/M_\Delta^2$ is greater than $10^{-3}~{\rm TeV}^{-2}\simeq 10^{-4} G_F.$
The former condition corresponds to
 neutrino masses greater than 0.05 eV  and neutrino phases which satisfy the relations $|\alpha_1+2\delta|,|\alpha_2+2\delta|\ll \pi.$

Equilibration between electron and non-electron species is a striking feature, absent in the standard picture of the collapse. It starts to settle not earlier than few milliseconds before bounce, when the density exceeds $10^{13}$ g/cm$^3.$  Such equilibration lifts the chemical potentials of muons and non-electron neutrinos higher than $100$ MeV, somewhat decreases the chemical potentials of electrons and electron neutrinos, leads to the increase of temperature and entropy. In addition to Fermi seas of extremely degenerate electrons and electron neutrinos, which are present in the standard picture, seas of extremely degenerate non-electron neutrinos and moderately degenerate muons emerge in case of LFV. This may affect the supernova physics in various ways. An elaborated study  should be carried out  in order to accurately and thoroughly describe all possible consequences. In particular, collapse simulations with new interactions included are necessary. In the present work we restrict ourselves by just outlining various  possibilities. They include modifications of supernova dynamics and modifications of supernova neutrino signal.

Possible modifications of supernova dynamics are basically due to the decrease of pressure in the center of the core, where LFV reactions proceed effectively. Such decrease gives rise to a non-homologous stage of the collapse which may start few ms before the ``standard'' bounce. This may lead to modification of the shock wave and even to additional shock waves. In addition, the increase of temperature and entropy may affect the delayed explosion scenario.

Possible modifications of supernova  neutrino signal include:
\begin{itemize}
\item Modification of the form, duration and total energy of the neutronization burst, which accompanies the shock breakout.
 If additional shock waves emerge due to LFV, then additional neutronization bursts are expected.
\item Modification of the total duration of the supernova neutrino signal, which signifies the alteration of the cooling rate of the proto-neutron star. There are three effects which affect the neutrino diffusion from the inner core: the emergence of seas of degenerate non-electron neutrinos, which do not take part in charge current interactions; the decrease of the average neutrino energy and the decrease of neutrino degeneracy. However, the first two effects tend to speed up the cooling, while the third one -- to slow down. In the present work we do not find which trend wins.
\end{itemize}

The discussed signatures in neutrino signal do not concern the flavor content of the neutrino signal and therefore are not smeared by neutrino oscillations.

We emphasize once more time that all effects listed above should be verified by profound supernova calculations. Once such calculations are done, it is possible to use the results in various ways. On the one hand, one can try to construct a successful supernova explosion model taking into account lepton flavor violation.   On the other hand, one can extract parameters of the see-saw type II model and other models which include  bileptons (or restrict such parameters) from the already persistent or future astrophysical data. In particular, one may get some bounds from the very fact that supernova explodes.

As a finale remark we note that the discussed signatures of LFV in supernova are rather general and may be relevant in the context of various microscopic models, not necessarily those which include bileptons.  In particular, the required LFV may be generated by the spin flips of Majorana neutrinos with large magnetic moment \cite{Lychkovskiy:2008da} or by
the flavor changing neutral currents  in  SUSY with R-parity violation \cite{Amanik:2004vm}.




\section*{Acknowledgments}

The authors thank N.V.~Mikheev, V.B.~Semikoz and A.Yu~Smirnov for useful discussions and acknowledge the partial support from grants NSh-2977.2008.2, NSh-3884.2008.2, NSh-4568.2008.2, RFBR-07-02-00830,
RFBR-07-02-00021, RFBR-08-02-00494 and from the RF Government contract No. 02.740.11.5158.
SB is grateful to K.Nomoto  and H.Mura\-yama for hospitality at  IPMU in Tokyo, his work
is also partly supported by World Premier International Research Center Initiative (WPI), MEXT,
Japan, by the Swiss National Science Foundation (SCOPES project No.~IZ73Z0-128180/1),
and by Federal Programm ``Science and innovation in Russia'', contract number
02.740.11.0250.
OL is grateful to the Dynasty Foundation for financial support.

\section*{Appendix: Comments on number of muon pairs in collapsing supernovae}
\label{sec:App}

The role of muons in collapsing supernovae was pointed out long
ago by Domogatsky \cite{domog}. He has shown that
for $T>20$ MeV the number density of muons is sufficient to trap
muon neutrinos  $\nu_\mu$  due to charge current reactions with muons only.
The role of neutral currents is understood much better nowadays,
and we know that $\nu_\mu$ and $\nu_\tau$ must be trapped
even without muons at lower
temperatures due to scattering on nuclei and electrons.
The role of muons is important in the models considered in this paper,
and also, e.g., for
leptonic photons discussed  by Okun \cite{okun}
(see also \cite{blinetal}, \cite{gnin} and references therein).

It is trivial to find the number of muon pairs for the case of standard physics following, e.g., Landau \& Lifshits \cite{lanlif}).
Here we add just a few methodological remarks (and point out one
place where Landau \& Lifshits make an incorrect statement).

The photons inside stars have the blackbody
Bose distribution, their chemical potential being equal to zero. The
muons and antimuons are in thermal equilibrium with the photons,
particularly, due to the reaction
\begin{equation}
{\mu}^-+{\mu}^+\rightleftharpoons {\rm photons} \; ,
\label{epairs}
\end{equation}
and this means that \cite{lanlif}
\begin{equation}
\mu_\mu \equiv \mu_{\mu^-} =-\mu_{\mu^+} \;,
\label{zvezda}
\end{equation}
where $\mu_{\mu^-}$ and $\mu_{\mu^+}$ are the chemical potentials for
${\rm \mu}^-$ and ${\rm \mu}^+$, respectively.

On the other hand, reactions
\begin{equation}
\begin{array}{c}
e^-+p \rightleftharpoons \nu_e + n \\
\mu^-+p \rightleftharpoons \nu_\mu + n
\end{array}
\end{equation}
ensure that
\be
\mu_\mu - \mu_{\nu_\mu}= \mu_e - \mu_{\nu_e}.
\ee
It may be shown that in the supernova core $\mu_{\nu_\mu}$ is negligible, which leads to
\be
\mu_\mu \simeq \mu_e - \mu_{\nu_e}.
\ee
Boom code simulations show that $\mu_\mu$ does not exceed 60 MeV.


The number of muons is
\begin{equation}
n_\mu = {\displaystyle
{\displaystyle g_0\int _{0}^{\infty }} {\displaystyle \frac {4\pi p^{2}\,}
{\exp((\sqrt{p^2+m_\mu^2}-\mu_\mu)/T) + 1}} \,\cdot\, {dp \over (2\pi)^{3}}}  \; ,
\label{npzeromu}
\end{equation}
$g_0=2$ being the statistical weight of
a muon.
We are interested in the conditions when the temperature is of order of a
few MeV or of a few tens
MeV, so muons are non-relativistic and non-degenerate, then one can use
the Boltzmann-like approximation:
\begin{equation}
n_\mu = {\displaystyle
{\displaystyle g_0\exp(-(m_\mu-\mu_\mu)/T)
\int _{0}^{\infty }} {\displaystyle \frac {4\pi p^{2}\,}
{\exp(p^2/2m_\mu T)}} \,\cdot\, {dp \over (2\pi)^{3}}}
 = g_0\exp(-(m_\mu-\mu_\mu)/T) \Biggl({m_\mu T \over 2\pi} \Biggr)^{3/2}\; .
\label{nzeromu}
\end{equation}

Let us put $\mu_\mu=0$ for a moment, then an exponential factor reads $\exp(-m_\mu/T)$. At the first glance it seems surprising that the number of pairs $n_\mu$
is expressed through $\exp(-m_\mu/T)$ and not through $\exp(-2m_\mu/T)$,
while the energy required to create a pair is $2m_\mu$. A simple explanation
of this paradox can be suggested using the Boltzmann distribution.
Let us denote the number of $\mu^+$ in the volume $V$ as $N^+$, then
indeed, relative to vacuum ($N_0=1$)
\begin{equation}
 {N^+ \over N_0} = {g^+ \over 1} \exp(-2m_\mu/T)
\label{boltz}
\end{equation}
where statistical weight $g^+$ must take into account
the contribution of all translational states of the muon $\mu^+$ in
the volume it occupies $V/N^+=1/n_\mu^+$.
The integral over $d^3p/(2\pi)^3$ with
the Boltzmann factor $\exp(-p^2/2m_\mu T)$ gives
$(m_\mu T /2\pi)^{3/2}$. The same factor is supplied to $g^+$
by each $\mu^-$ born
together with the $\mu^+$ somewhere in the volume $V$, so this must be
multiplied by the whole $V$ now, and we find
\begin{equation}
 g^+=g_0^2 \Biggl({m_\mu T \over 2\pi} \Biggr)^3 {1 \over n_\mu^+} V \; .
\label{weight}
\end{equation}
Since $N^+=N^-$ we recover $n_\mu^+ n_\mu^-$ found above in
(\ref{nzeromu}).

It is interesting to compare the value of $n_\mu$ from (\ref{nzeromu})
with the electron concentration $n_e$.
We must notice that in the book by Landau \& Lifshits \cite{lanlif} \S 105
it is written not correctly that for
$T >m_e$ the number of electron-positron pairs
is high. It is true
only for the ``terrestrial'' densities. Inside the stars the situations
when $T>m_e$, but $T<\mu_e$, are not rare.
The equilibrium condition for electrons does not imply the equality
of the electron and positron concentrations: when the temperature
is lower than the chemical potential $\mu_e$,
the number of the positrons
is exponentially lower than that of the electrons by a factor of
$\exp(-2\mu_e /T)$. During the collapse,
the value of density is like $\rho=2\times 10^{14}$ \mbox{g/cm$^3$} and
for $Y_e=0.3$ this
gives $\mu_e=200$ MeV in eq.~(\ref{mue}), so the number of positrons is small even for
$T$ of order tens MeV. The correct expression for electrons is
\begin{equation}
n_e={\displaystyle \frac {g_e}{6\pi^2}} \,
   {\displaystyle {\mu_e^3} } \; .
\label{ne}
\end{equation}
Thus in the standard scenario
\be
n_\mu/ n_e = \frac{3\sqrt{\pi}}{\sqrt{2}} \exp(-(m_\mu-\mu_\mu)/T)\frac{(m_\mu T)^{3/2}}{\mu_e^3} \ll 1
\ee
in the supernova core.
We see that LFV dramatically raises the number density of muons making $n_\mu \sim n_e$.

\end{document}